\documentclass[onecollarge]{svjour2}       
\smartqed  
\usepackage{graphicx}
\usepackage{amsmath}
\journalname{Foundations of Physics}
\begin{document}

\title{An Inquiry into the Possibility of Nonlocal Quantum Communication
}


\author{John G. Cramer         \and
        Nick Herbert 
}


\institute{John G. Cramer \at
              Dept. of Physics, Box 351560, Univ. of Washington, Seattle WA 98195-1560\\
              Tel.: +001 206-525-3504\\
              \email{jcramer@uw.edu}           
           \and
           Nick Herbert \at
              Box 261, Boulder Creek, CA 95006
}

\date{Received: 16 September, 2014 / Revised: 14 February, 2015 / Accepted: date}

\maketitle

\begin{abstract}
The possibility of nonlocal quantum communication is considered.  We investigate three \textit{gedankenexperiments} that have variable entanglement: (1) a 4-detector polarization-entangled system, (2) a 4-detector path-entangled system, and (3) a 3-detector path-entangled system that uses an innovative optical mixer to combine photon paths.  A new quantum paradox is reviewed in which the presence or absence of an interference pattern in a path-entangled two photon system, controlled by measurement choice, is a potential nonlocal signal.  We show that for the cases considered, even when interference patterns can be switched off and on, there is always a ``signal" interference pattern and an ``anti-signal" interference pattern that mask any observable interference when they are added, even when entanglement and coherence are simultaneously present.  This behavior can be attributed to what in the literature has been called ``the complementarity of one- and two-particle interference". 
\keywords{quantum \and nonlocal \and communication \and interference \and complementarity \and entanglement}
\PACS{03.65.Aa \and 03.65.Ud \and 03.67.Hk}
\end{abstract}

\section{Introduction}
\label{sec:1}
Quantum mechanics, our standard model of the physical world at the smallest scales of energy and size, has built-in retrocausal aspects.  For example, Wheeler's delayed choice \textit{gedankenexperiment}\cite{Wh78} describes a scheme in which the experimenter's later choice of measurement retroactively determines whether a light photon that had previously encountered a two-slit aperture had passed through both slits or through only one slit.

In the present work we describe a new quantum-mechanical paradox in which the presence or absence of an interference pattern in a path-entangled two photon system with variable entanglement, controlled by measurement choice, would seem to permit retrocausal signaling from one observer to another.  We also present an analysis of this scheme, showing how the subtleties of the quantum formalism block the potential signal.   In particular, even when interference patterns can be switched off and on, there is always a ``signal" interference pattern and an ``anti-signal" interference pattern that mask any observable interference when they are added, even when entanglement and coherence are simultaneously present.  This behavior can be attributed to what in the literature has been called ``the complementarity of one- and two-particle interference"\cite{Ja93}.

\section{Quantum Nonlocality and Entanglement}
\label{sec:2}
Quantum mechanics differs from the classical mechanics of Newton that preceded it in one very important way.  Newtonian systems are always \textit{local}.  If a Newtonian system breaks up, each of its parts receives a definite and well-defined energy, momentum, and angular momentum, parceled out at breakup by the system while respecting the conservation laws. After the component subsystems are separated, the properties of each subsystem are completely independent and do not depend on those of the other subsystems.
 
On the other hand, quantum mechanics is \textit{nonlocal}, meaning that the component parts of a quantum system may continue to influence each other, even when they are well separated in space and out of speed-of-light contact.  This unexpected characteristic of standard quantum theory was first pointed out by Albert Einstein and his colleagues Boris Podolsky and Nathan Rosen (EPR) in 1935, in a critical paper\cite{Ei35} in which they held up the discovered nonlocality as a devastating flaw that, it was claimed, demonstrated that the standard quantum formalism must be incomplete or wrong.  Einstein called nonlocality ``spooky actions at a distance".  Schr\"{o}dinger followed on the discovery of quantum nonlocality by showing in detail how the components of a multi-part quantum system must depend on each other, even when they are well separated\cite{Sc35}.

Beginning in 1972 with the work of Stuart Freedman and John Clauser\cite{Fr72}, a series of quantum-optics EPR experiments testing Bell-inequality violations\cite{Be66} and other aspects of linked quantum systems were performed.  These experimental results can be taken as demonstrating that, like it or not, both quantum mechanics and the reality it describes are intrinsically nonlocal.  Einstein's spooky actions-at-a-distance are really out there in the physical world, whether we understand and accept them or not.

How and why is quantum mechanics nonlocal?  Nonlocality comes from two seemingly conflicting aspects of the quantum formalism: (1) energy, momentum, and angular momentum, important properties of light and matter, are conserved in all quantum systems, in the sense that, in the absence of external forces and torques, their net values must remain unchanged as the system evolves, while (2) in the wave functions describing quantum systems, as required by the uncertainty principle, the conserved quantities are often indefinite and unspecified and typically can span a large range of possible values. This non-specificity persists until a measurement is made that ``collapses" the wave function and fixes the measured quantities with specific values. These seemingly inconsistent requirements of (1) and (2) raise an important question: how can the wave functions describing the separated members of a system of particles, which may be light-years apart, have arbitrary and unspecified values for the conserved quantities and yet respect the conservation laws when the wave functions are collapsed?

This paradox is accommodated in the formalism of quantum mechanics because the quantum wave functions of particles are \textit{entangled}, the term coined by Schr\"{o}dinger to mean that even when the wave functions describe system parts that are spatially separated and out of light-speed contact, the separate wave functions continue to depend on each other and cannot be separately specified.  In particular, the conserved quantities in the system's parts (even though individually indefinite) must always add up to the values possessed by the overall quantum system before it separated into parts.

How could this entanglement and preservation of conservation laws possibly be arranged by Nature?  The mathematics of quantum mechanics gives us no answers to this question, it only insists that the wave functions of separated parts of a quantum system do depend on each other.  Theorists prone to abstraction have found it convenient to abandon the three-dimensional universe and describe such quantum systems as residing in a many-dimensional Hilbert hyper-space in which the conserved variables form extra dimensions and in which the interconnections between particle wave functions are represented as allowed sub-regions of the overall hyper-space.  That has led to elegant mathematics, but it provides little assistance in visualizing what is really going on in the physical world.

In this paper, for reasons of space and focus, we will not attempt to account for nonlocality by considering any interpretation of quantum mechanics.  We will simply note that the transactional interpretation\cite{Cr86} of quantum mechanics, introduced by one of the authors in 1986, seems to be unique among the plethora of interpretations of the quantum formalism in providing a definite mechanism that accounts for nonlocality and facilitates visualization of nonlocal processes.  Here we will take the existence of quantum nonlocality and entanglement as established facts and consider their implications.

\section{No-Signal Theorems}
\label{sec:3}

Given that a measurement on one part of an extended quantum system can affect the outcomes of measurements performed in other distant parts of the system, the question that naturally arises is: \textit{can this phenomenon be used for nonlocal communication between one observer and another?}  Demonstration of such nonlocal quantum communication would be a truly ``game-changing" discovery, because it would break all the rules of normal communication.  No energy would pass between the send and receive stations; the acts of sending and receiving could occur in either time order and would depend only on the chosen instants at which the measurements were made; there would be no definite signal-propagation speed, and messages could effectively be sent faster than light-speed, or ``instantaneously" in any chosen reference frame, or even, in principle, backwards in time.

The average member of the physics community, if he or she has any opinion about nonlocal communication at all, believes it to be impossible, in part because of its superluminal and retrocausal implications.  Over the years a number of authors have presented proofs, based on the standard quantum formalism, showing that nonlocal observer-to-observer communication is impossible\cite{Eb77,Gh80,Yu05}. They employ details of quantum mechanics and quantum field theory and show that in separated measurements involving entangled quantum systems, the quantum correlations will be preserved but there will be no effect apparent to an observer in one sub-system if the character of the measurement is changed in the other sub-system. Thus, the standard quantum formalism implies that nonlocal signaling is impossible, and any hypothetical observation of nonlocal signaling would require some change in that formalism.

Nevertheless, it is interesting to observe in specific seemingly paradoxical cases (see below) exactly how the nonlocal signal is blocked.  We note that one plausible mechanism for blocking nonlocal signals is the known ``complementary" or see-saw relation between entanglement and coherence\cite{Ab01}.  Since some degree of \textit{both} coherence and entanglement would be required for any potential nonlocal signal\cite{Cr09}, it is particularly interesting to study systems in which the entanglement/coherence ratio is a parameter that can be varied so that both are simultaneously present (see below), as would be reqired for an interference ``signal".\\

We note that it is also sometimes asserted that nonlocal communication is not possible because it would conflict with special relativity. This assertion is incorrect. The prohibition of signals with superluminal speeds by Einstein's theory of special relativity is related to the fact that a condition of definite simultaneity between two separated space-time points is not Lorenz invariant.  Assuming that some hypothetical superluminal signal could be used to establish a fixed simultaneity relation between two such points, e.g., by clock synchronization, this would imply a preferred inertial frame and would be inconsistent with Lorenz invariance and special relativity. In other words, superluminal signaling would be inconsistent with the even-handed treatment of all inertial reference frames that is the basis of special relativity. 

However, if a nonlocal signal could be transmitted through measurements at separated locations performed on two entangled photons, the signal would be ``sent" at the time of the arrival of one photon at one location and ``received" at the time of arrival of the other photon at the other location, both along Lorenz-invariant light-like world lines.  By varying path lengths to the two locations, these events could be made to occur in any order and time separation in any reference frame.  Therefore, nonlocal signals (even superluminal and retrocausal ones) could not be used to establish a fixed simultaneity relation between two separated space-time points, because the sending and receiving of such signals do not have fixed time relations.  Nonlocal quantum signaling, if it were to exist, would be completely compatible with special relativity.  (However, it would probably \textit{not} be compatible with macroscopic causality.)

\section{A Polarization-entangled EPR Experiment with Varaible Entanglement}
\label{sec:4}

First, let us examine a fairly simple EPR experiment exhibiting nonlocality.  Following Bell\cite{Be66}, a number of experimental EPR tests\cite{Fr72,As82} have exploited the correlations of polarization-entangled systems that arise from angular momentum conservation.  Their results, to accuracies of many standard deviations, are consistent with the predictions of standard quantum mechanics and can be interpreted as falsifying many local hidden-variable alternatives to quantum mechanics.

\begin{figure}
 \includegraphics[width=15 cm]{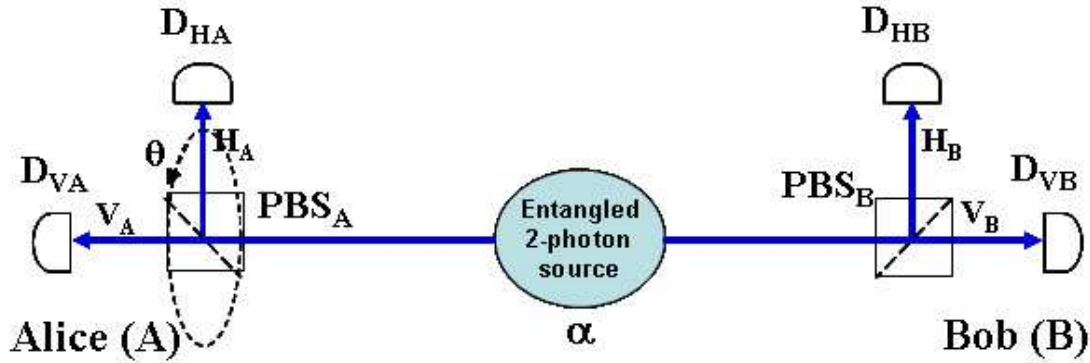}
\caption{(color online) A two-photon 4-detector EPR experiment using linear polarization with variable entanglement.}
\label{fig:1}       
\end{figure}

A modern version of this type of EPR experiment, one in which the entanglement/coherence ratio is an adjustable parameter, is shown in Fig. \ref{fig:1}.  We note that while there are many analyses of such experiments in the literature, there is no previous analysis for a system with variable entanglement.

Two observers, Alice and Bob, operate polarimeters measuring the linear polarization (H or V) of individual photons and record photon detections.  The H-V plane of Alice's polarimeter can be rotated through an angle $\theta$ with respect to the plane of Bob's polarimeter, so that the basis of her polarization measurements can be changed relative to Bob's.

Here the source of photons is taken to be a Sagnac entangled two-photon source of the type developed by the Zeilinger Group\cite{Fe07}, in which the degree of entanglement can be varied by rotating a half-wave plate in the system, as characterized by the variable $\alpha$, producing a two-particle wave function of the general form:\\
$\Psi(\alpha)= (\mid\!H_{A}\rangle\mid\!H_{B}\rangle+\mid\!V_{A}\rangle\mid\!V_{B}\rangle)(\cos\beta+\sin\beta)/2\\
~~~~~~~+ i(\mid\!H_{A}\rangle\mid\!V_{B}\rangle-\mid\!V_{A}\rangle\mid\!H_{B}\rangle)(\cos\beta-\sin\beta)/2$ where $\beta=\alpha-\pi/4$.  The degree of photon-pair entanglement from this source is adjustable.  When $\alpha=0$, the two-photon polarization entanglement is 100\% in a pure Bell state with the wave function $\Psi(0) = i(\mid\!H_{A}\rangle\mid\!V_{B}\rangle-\mid\!V_{A}\rangle\mid\!H_{B}\rangle)/\sqrt{2}$; when $\alpha=\pi/4$ the entanglement is 0 in a non-entangled product state with $\Psi(\pi/4) = [(\mid\!H_{A}\rangle-i\mid\!V_{A}\rangle)\times(\mid\!H_{B}\rangle+i\mid\!V_{B}\rangle)]/2$; when $\alpha=\pi/8$ the source will produce photon pairs with 71\% entanglement and 71\% coherence.
 
\begin{figure}
 \includegraphics[width=15 cm]{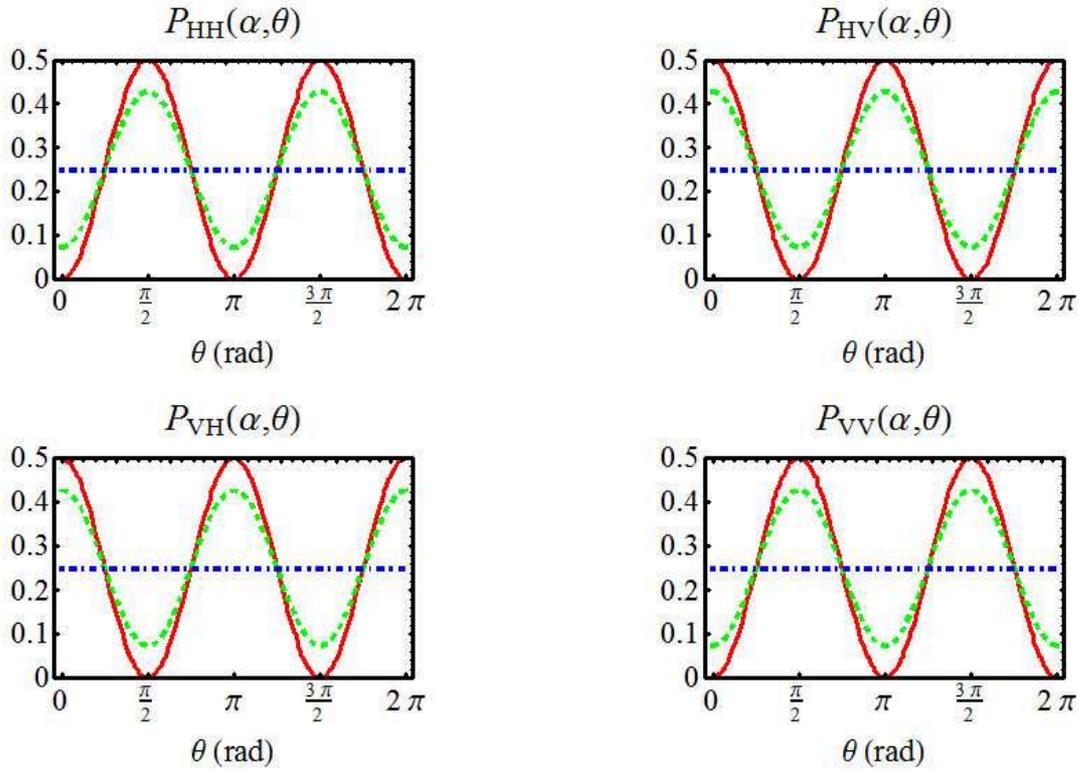}
\caption{(color online) Joint detection probabilities vs. $\theta$ for the four detector combinations with: $\alpha=0$  (red/solid, 100\% entangled), $\alpha=\pi/8$  (green/dashed, 71\% entangled), and $\alpha=\pi/4$  (blue/dot-dashed, 0\% entangled)}
\label{fig:2}       
\end{figure}

Let us assume that we initially set $\alpha=0$ for 100\% entanglement.  When $\theta$ is zero and the polarimeters are aligned, there will be a perfect anti-correlation between the polarizations measured by Alice and by Bob.  The random polarization (H or V) that Alice measures will always be the opposite of that measured by Bob ($H_{A}V_{B}$ or $V_{A}H_{B}$).  However, when $\theta$ is increased, the perfect $H_{A}V_{B}$ and $V_{A}H_{B}$ anti-correlations are degraded and correlated detections $H_{A}H_{B}$ and $V_{A}V_{B}$, previously not present, will begin to appear.

Local theories require that for small $\theta$ rotations this correlation degradation should increase linearly with $\theta$, while quantum mechanics predicts that it should increase as $\theta^{2}$, i.e., quadratically\cite{He75}.  This is the basis of Bell's Inequalities\cite{Be66}, counting ratio inequalities that are valid for linear behavior in $\theta$ but are dramatically violated for the quadratic behavior characteristic of quantum mechanics.

The quantum mechanical analysis of this system is fairly simple because, assuming that the entangled photons have a single spatial mode, their transport through the system can be described by considering only the phase shifts and polarization selections that the system elements create in the waves.  We have used the formalism of Horne, Shimony and Zeilinger\cite{Ho90} to perform such an analysis and to calculate the joint wave functions for simultaneous detections at both detectors\cite{Ma14a}.  These are:
\begin{align}
	\Psi_{HH}(\alpha,\theta) = [-\sin(\alpha)\cos(\theta)+i\cos(\alpha)\sin(\theta)]/\sqrt{2}\\
	\Psi_{HV}(\alpha,\theta) = [-\cos(\alpha)\cos(\theta)+i\sin(\alpha)\sin(\theta)]/\sqrt{2}\\
	\Psi_{VH}(\alpha,\theta) = [\cos(\alpha)\cos(\theta)-i\cos(\alpha)\sin(\theta)]/\sqrt{2}\\
	\Psi_{VV}(\alpha,\theta) = [\sin(\alpha)\cos(\theta)-i\cos(\alpha)\sin(\theta)]/\sqrt{2}.
\end{align}
The corresponding joint detection probabilities are: 
\begin{align}
        P_{HH}(\alpha,\theta) = [1 - \cos(2\alpha)\cos(2\theta)]/4\\
        P_{HV}(\alpha,\theta) = [1 + \cos(2\alpha)\cos(2\theta)]/4\\
	P_{VH}(\alpha,\theta) = [1 + \cos(2\alpha)\cos(2\theta)]/4\\
	P_{VV}(\alpha,\theta) = [1 - \cos(2\alpha)\cos(2\theta)]/4.
\end{align}
Fig. \ref{fig:2} shows plots of these joint detection probabilities vs. $\theta$ for the four detector combinations with: $\alpha=0$  (100\% entangled), $\alpha=\pi/8$  (71\% entangled), and $\alpha=\pi/4$  (0\% entangled) .

Now consider the question of whether, at any setting of $\alpha$, observer Alice by operating the left system and varying $\theta$ can send a nonlocal signal to observer Bob operating the right system.  Some overall observer who is monitoring the coincidence counting rates $H_{A}H_{B}$, $V_{A}V_{B}$, $H_{A}V_{B}$, and $V_{A}H_{B}$ could reproduce Fig. \ref{fig:2} and would have a clear indication of when $\theta$ was varied by Alice, in that the relative rates would change dramatically.  However, observer Bob is isolated at the system on the right and is monitoring only the two singles counting rates $H_{B} \equiv H_{A}H_{B}+V_{A}H_{B}$ and $V_{B} \equiv H_{A}V_{B}+V_{A}V_{B}$.

Bob would observe the probabilities $P_{BH}(\alpha,\theta) = P_{HH}(\alpha,\theta) + P_{VH}(\alpha,\theta) =\frac{1}{2}$ and  $P_{BV}(\alpha,\theta) = P_{HV}(\alpha,\theta) + P_{VV}(\alpha,\theta) =\frac{1}{2}$, both independent of the values of $\alpha$ and $\theta$.  Thus, Bob would see only counts detected at random in one or the other of his detectors with a 50\% chance of each polarization, and his observed rates would not be affected by the setting of $\theta$.  Alice's choice of her $\theta$ setting will alter the wave functions that arrive at Bob's detectors, but not in a way that permits signaling.   Schr\"{o}dinger called this effect ``steering" the wave functions\cite{Sc35}.  The late Heinz Pagels, in his book \textit{The Cosmic Code}\cite{Pa82}, examined in great detail the way in which the intrinsic randomness of quantum mechanics blocks any potential nonlocal signal in this type of polarization-based EPR experiment.

We emphasize the point that linear polarization is an interference effect of the photon's intrinsic circularly-polarized spin angular momentum $ S=1$, $S_z=\pm 1$ helicity eigenstates.  As we will see below, the interference blocking observed here is an example of  a ``signal" interference pattern and an ``anti-signal" interference pattern that mask any observable interference when they are added, even when entanglement and coherence are simultaneously present.  This behavior is attributed to what has been called the complementarity of one-particle and two-particle interference\cite{Ja93}.

\section{A Path-entangled EPR Experiment with Variable Entanglement}
\label{sec:5}

Although the entanglement of linear polarization is a very convenient medium for EPR experiments and Bell-inequality tests, in many ways the alternative offered by path-entangled EPR experiments provides a richer venue.  Perhaps the earliest example of a path-entangled EPR experiment is the 1995 ``ghost interference" experiment of the Shih Group at University of Maryland Baltimore County\cite{St95}.  Their experiment demonstrated that an interference pattern observed for one member of a pair of entangled photons could be ``switched" off or on depending on whether the other photon of the pair went through one slit or both slits of a two-slit aperture.

Another path-entangled EPR experiment was the 1999 PhD thesis of Dr. Birgit Dopfer at the University of Innsbruck\cite{Do99}, performed under the direction of Prof. Anton Zeilinger.  This experiment demonstrated that one could can make the interference pattern observed for one of a pair of entangled photons appear or disappear, depending on whether the location of the detector that detected the other member of the entangled pair was at or away from the focal point of a lens.

Examination of the ghost-interference and Dopfer experiments raises a very interesting question: Can the requirement of a coincidence  between the entangled photons, used in both experiments,  be removed?  The answer to this question is subtle.  In principle, the two entangled photons are connected by nonlocality whether they are detected in coincidence or not, so the coincidence may perhaps be removable.  However, in both experiments the authors reported that no two-slit interference distribution was observed when the coincidence requirement was removed.  These considerations lead to a new quantum mechanical paradox: they suggest\cite{Je06} that if the coincidence requirement could be relaxed, nonlocal observer-to-observer signals might be transmitted by controlling the presence or absence of an interference pattern, essentially by forcing wave-like or particle-like behavior on both members of an entangled photon pair.

From the point of view of moving to a path-entanglement situation in which the coincidence requirement could be relaxed, the problem with both of the experiments discussed above is their use of a two-slit system that blocks and absorbs most of the photons from the nonlinear crystal that illuminate the slit system.  The down-conversion process is intrinsically very inefficient ($\sim$1 photon pair per $10^{8}$ pump photons), so there are no photons to waste.  An additional complication is that most detectors capable of detecting individual photons are intrinsically noisy and somewhat inefficient.  For these reasons, there is a large advantage in using \textit{all} of the available entangled-photon pairs in any contemplated path-interference test of nonlocal communication.
  
\begin{figure}
 \includegraphics[width=15 cm]{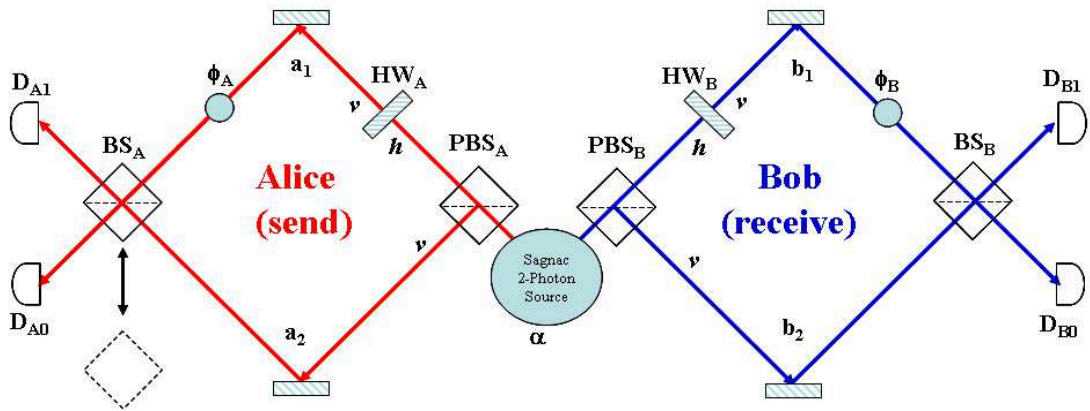}
\caption{(color online) A 4-detector path-entangled dual-interferometer EPR experiment with variable entanglement.}
\label{fig:3}       
\end{figure}

Fig. \ref{fig:3} shows a path-entangled experimental EPR test using two Mach-Zehnder interferometers\cite{Ze91} that have been modified to convert polarization entanglement to path entanglement\cite{Zu88}.  This type of system was originally developed by the Zeilinger Group at the Institute for Quantum Optics and Quantum Information, Vienna\cite{Fe08}.  Here, the interferometers are a variant of the basic Mach-Zehnder design that uses an initial polarizing beam splitter ($PBS_{A,B}$) that directs the vertical ($v$) and horizontal ($h$) linear polarizations to different paths and then converts horizontal to vertical polarization on the upper path with a half-wave plate ($HW_{A,B}$).  This has the effect of converting polarization entanglement from the source to path entanglement and then placing waves on both paths in the same polarization state, so that they can interfere. Again observers Alice and Bob operate the interferometers and count and record individual photon detections.  A phase shift element ($\phi_{A,B}$) allows the observers to alter the phase of waves on the upper paths.

As in the polarization-entangled EPR example, the source of photons is taken to be the Sagnac entangled two-photon source developed by the Zeilinger Group\cite{Fe07}, in which the degree of entanglement depends on the value of $\alpha$ in this setup.  Following the path separation the extended two-particle wave function has the general form:\\
$\Psi(\alpha) = (\mid\!a_1\rangle\mid\!b_1\rangle+\mid\!a_2\rangle\mid\!b_2\rangle)(\cos\beta+\sin\beta)/2\\
~~~~~~+ i(\mid\!a_1\rangle\mid\!b_2\rangle-\mid\!a_2\rangle\mid\!b_1\rangle)(\cos\beta-\sin\beta)/2$ where $\beta=\alpha-\pi/4$.  When $\alpha=\pi/2$, the two-photon wave function is a fully path-entangled Bell state of the form $\Psi(\pi/2) = (\mid\!a_1\rangle\mid\!b_1\rangle+\mid\!a_2\rangle\mid\!b_2\rangle)/\sqrt{2}$, and when $\alpha=\pi/4$ the path entanglement is 0 and the wave function is a product state of the form $\Psi(\pi/4) = (\mid\!a_1\rangle-i\mid\!a_2\rangle)\times(\mid\!b_1\rangle+i\mid\!b_2\rangle)/2$.

Alice's last beam-splitter ($BS_{A}$) is removable.  When $BS_{A}$ is in place, the two left paths are remixed, the left-going photons exhibit the wave-like behavior of being on both paths, and two-path overlap and Mach-Zehnder interference will be present.  When $BS_A$ is removed, path detection occurs, the left-going photons exhibit the particle-like behavior of being on a path uniquely ending at detector $D_{A0}$ or at detector $D_{A1}$, so that Alice's measurements provide which-way information about both photons.  Bob's last beam splitter ($BS_{B}$) remains in place and, in the absence of which-way information, should exhibit Mach-Zehnder interference.

This experiment is thus the equivalent of the ghost-interference experiment and the Dopfer experiment described above, in that it embodies entangled paths and two-path interference.  However, it improves on those experiments by using all of the available entangled photons and by employing a source that has a variable entanglement that depends on $\alpha$.

It has been argued\cite{Je06,Cr09} that this situation presents a nonlocal signaling paradox, in that Alice. by choosing whether $BS_{A}$ is in or out, can cause the Mach-Zehnder interference effect to be present or absent in Bob's detectors.  In particular, with $BS_{A}$ out we expect particle-like behavior, and Bob should observe equal counting rates in $D_{B1}$ and $D_{B0}$.  With $BS_{A}$ in we expect wave-like behavior, and  Bob, for the proper choice of $\phi_{B}$, should observe all counts in $D_{B1}$ and no counts in $D_{B0}$ due to Mach-Zehnder interference.  It was further argued\cite{Cr09} that possibly the nonlocal signal might be suppressed by the complementarity of entanglement and coherence\cite{Ab01}, and that by arranging for 71\% entanglement and 71\% coherence (i.e., $\alpha=\pi/8$ for the Sagnac source), a nonlocal signal might be permitted.

As in the EPR example discussed above, the quantum mechanical analysis of this system is fairly simple because, assuming that the entangled photons have a single spatial mode, their transport through the system can be described by considering the phase shifts that the system elements create in the waves.  To test the validity of the above arguments, we have used the formalism of Horne, Shimony and Zeilinger\cite{Ho90} in \textit{Mathematica 9} to analyze the dual-interferometer configuration\cite{Ma14b} and to calculate the joint wave functions for detections of the entangled photon pairs in various combinations.

For $BS_{A}$ in, these wave functions are:
\begin{align}
\Phi_{A_1B_1}&(\alpha,\phi_A,\phi_B) = [i\cos(\alpha)(e^{i\phi_A}-e^{i\phi_B})\nonumber\\
&+\sin(\alpha)(1+e^{i(\phi_A+\phi_B)})]/(2\sqrt{2})\\
\Phi_{A_1B_0}&(\alpha,\phi_A,\phi_B) = [-\cos(\alpha)(e^{i\phi_A}+e^{i\phi_B})\nonumber\\
&+i\sin(\alpha)(1-e^{i(\phi_A+\phi_B)})]/(2\sqrt{2})\\
\Phi_{A_0B_1}&(\alpha,\phi_A,\phi_B) = [\cos(\alpha)(e^{i\phi_A}+e^{i\phi_B})\nonumber\\
&+i\sin(\alpha)(1-e^{i(\phi_A+\phi_B)}]/(2\sqrt{2})\\
\Phi_{A_0B_0}&(\alpha,\phi_A,\phi_B) = [i\cos(\alpha)(e^{i\phi_A}-e^{i\phi_B})\nonumber\\
&-\sin(\alpha)(1+e^{i(\phi_A+\phi_B)}]/(2\sqrt{2}).
\end{align}

The corresponding joint detection probabilities are:
\begin{align}
P_{A_1B_1}&(\alpha,\phi_A,\phi_B) = \{1-\sin(\phi_A)[\sin(2\alpha)+\sin(\phi_B)]\nonumber\\
&-\cos(2\alpha)\cos(\phi_A)\cos(\phi_B)+\sin(2\alpha)\sin(\phi_B)\}/4\\
P_{A_1B_0}&(\alpha,\phi_A,\phi_B) = \{1-\sin(2\alpha)[(\sin(\phi_A)+\sin(\phi_B)]\nonumber\\
&+\cos(2\alpha)\cos(\phi_A)\cos(\phi_B)+\sin(2\alpha)\sin(\phi_B)\}/4\\
P_{A_0B_1}&(\alpha,\phi_A,\phi_B) = \{1+\sin(2\alpha)[(\sin(\phi_A)+\sin(\phi_B)]\nonumber\\
&+\cos(2\alpha)\cos(\phi_A)\cos(\phi_B)+\sin(2\alpha)\sin(\phi_B)\}/4\\
P_{A_0B_0}&(\alpha,\phi_A,\phi_B) = \{1-\sin(\phi_B)[\sin(2\alpha)+\sin(\phi_A)]\nonumber\\
&-\cos(2\alpha)\cos(\phi_A)\cos(\phi_B)+\sin(2\alpha)\sin(\phi_A)\}/4.
\end{align}

The non-coincident singles detector probabilities for Bob's detectors are obtained by summing over Alice's detectors, which he does not observe.  Thus
\begin{align}
\label{eq:B1}
P_{B_1}(\alpha,\phi_B)& \equiv P_{A_1B_1}(\alpha,\phi_A,\phi_B) + P_{A_0B_1}(\alpha,\phi_A,\phi_B)\nonumber\\
& = [1 + \sin(2\alpha)\sin(\phi_B)]/2\\
\label{eq:B0}
P_{B_0}(\alpha,\phi_B)& \equiv P_{A_1B_0}(\alpha,\phi_A,\phi_B) + P_{A_0B_0}(\alpha,\phi_A,\phi_B)\nonumber\\
&= [1 - \sin(2\alpha)\sin(\phi_B)]/2.
\end{align}
Note that these singles probabilities have no dependences on Alice's phase $\phi_A$ for any value of $\alpha$.  Here again we see an example of  Schr\"{o}dinger steering, in that Alice is manipulating the wave functions that arrive at Bob's detectors, but not in such a way that would permit signaling.

\begin{figure}
\includegraphics[width=15 cm]{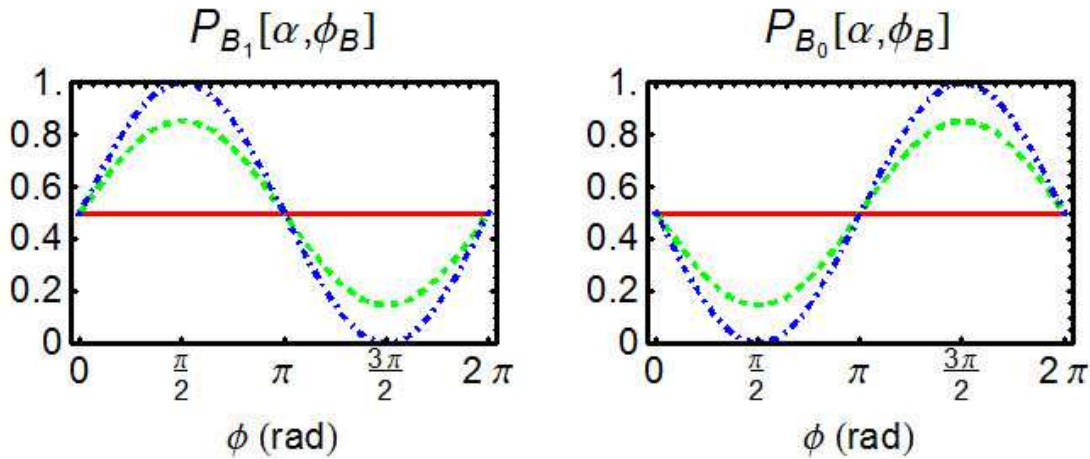}
\caption{(color online) Bob's non-coincident singles detector probabilities $P_{B_1}(\alpha,\phi_B)$ and $P_{B_0}(\alpha,\phi_B)$ (Eqns. \ref{eq:B1} and \ref{eq:B0}) for $\alpha=0$ (red/solid, 100\% entangled), $\alpha=\pi/8$ (green/dash, 71\% entangled), and $\alpha=\pi/4$ (blue/dot-dash, 0\% entangled).
}
\label{fig:4}       
\end{figure}

Fig. \ref{fig:4} shows plots of Bob's non-coincident singles detector probabilities $P_{B_1}(\alpha,\phi_B)$ and $P_{B_0}(\alpha,\phi_B)$ for the cases of $\alpha=0$ (100\% entangled), $\alpha=\pi/8$ (71\% entangled), and $\alpha=\pi/4$ (not  entangled).
 
We see here a demonstration of the see-saw relation between entanglement and coherence\cite{Ab01}, in that the probabilities for fully entangled system are constant, independent of $\phi_{B}$, because the absence of coherence suppresses the Mach-Zehnder interference, while the unentangled system shows strong Mach-Zehnder interference.  The $\alpha=\pi/8$ case, with 71\% coherence and entanglement, also shows fairly strong Mach-Zehnder interference and raises the intriguing possibility that a nonlocal signal might survive.

Therefore, the question raised by the possibility of nonlocal signaling is: \textit{What happens to Bob's detection probabilities when Alice's beam splitter $BS_{A}$ is removed?}  To answer this question, we re-analyze the dual interferometer experiment of Fig. \ref{fig:6} with $BS_{A}$ in the ``out" position.  These calculations\cite{Ma14c} give the joint wave functions for simultaneous detections of detector pairs:
\begin{align}
\Psi_{A_1B_1}(\alpha,\phi_A,\phi_B) = [\sin(\alpha)-i e^{i\phi_B}\cos(\alpha)/2\\
\Psi_{A_1B_0}(\alpha,\phi_A,\phi_B) = [i\sin(\alpha)-e^{i\phi_B}\cos(\alpha)]/2\\
\Psi_{A_0B_1}(\alpha,\phi_A,\phi_B) = [e^{i\phi_A}(\cos(\alpha)-ie^{i\phi_B}\sin(\alpha)]/2\\
\Psi_{A_0B_0}(\alpha,\phi_A,\phi_B) = [i e^{i\phi_A}(\cos(\alpha)+i e^{i\phi_B}\sin(\alpha)]/2.
\end{align}

The corresponding joint detection probabilities are: 
\begin{align}
P_{A_1B_1}(\alpha,\phi_A,\phi_B) = [1 + \sin(2\alpha)\sin(\phi_B)]/4\\
P_{A_1B_0}(\alpha,\phi_A,\phi_B) = [1 - \sin(2\alpha)\sin(\phi_B)]/4\\
P_{A_0B_1}(\alpha,\phi_A,\phi_B) = [1 + \sin(2\alpha)\sin(\phi_B)]/4\\
P_{A_0B_0}(\alpha,\phi_A,\phi_B) = [1 - \sin(2\alpha)\sin(\phi_B)]/4.
\end{align}
The non-coincident singles detector probabilities for Bob's detectors are identical to the singles detector probabilities of Eqns. \ref{eq:B1} and \ref{eq:B0} obtained when $BS_{A}$ was in place.

The conclusion is that, for any value of $\alpha$, no nonlocal signal can be sent by inserting and removing $BS_{A}$ or by varying phase $\phi_{A}$.  We have also found (not shown here) that even when the left-going photons from the source are intercepted \textit{before} entering Alice's interferometer with a black absorber, Bob will still observe the same singles counting rates given by Eqns. \ref{eq:B1} and \ref{eq:B0}.  As in the polarization-entangled EPR case, the interference blocking observed is an example of  a ``signal" interference pattern and an ``anti-signal" interference pattern that mask any observable interference when they are added, even when entanglement and coherence are simultaneously present.  This behavior is again attributed to what has been called the complementarity of one-particle and two-particle interference\cite{Ja93}.

\section{A Wedge-modified Path-entangled EPR Experiment with Variable Entanglement}
\label{sec:6}

\begin{figure}
\includegraphics[width=15 cm]{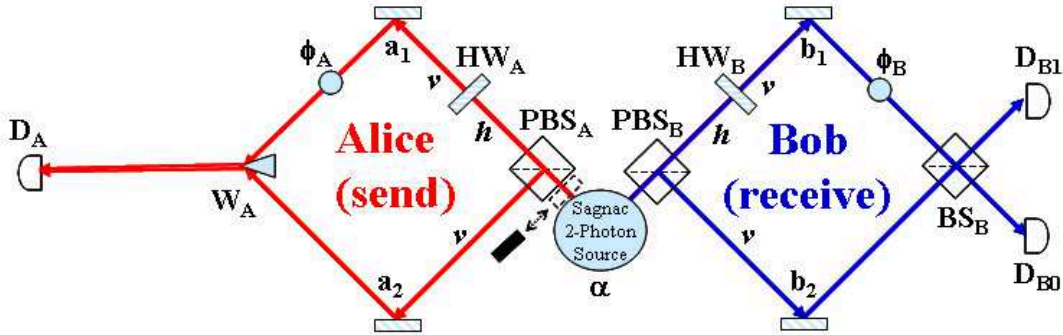}
\caption{(color online) A 3-detector wedge modification of the path-entangled dual-interferometer EPR experiment with variable entanglement.
}
\label{fig:5}       
\end{figure}

A possible reason that all of the above attempts at nonlocal communication have failed is that the left-going photons are directed to both of Alice's detectors.  The two detectors measure complementary interference profiles, so that when these profiles are added the potential nonlocal signal is erased.  Suppose that instead we direct all the photons on both paths to a single detector, where they should have only one interference profile.  Could this change permit nonlocal signaling?  To investigate this question we have analyzed the experiment shown in Fig. \ref{fig:5}.
 
Here, we have replaced Alice's last beam splitter and detectors with a somewhat unorthodox optical device, a $45^\circ$ wedge mirror $W_A$ that directs the left-going photons on paths $a_{1}$ and $a_{2}$ to a single detector $D_{A}$.  We assume that the angles of Alice's mirrors are tweaked slightly so that the two beams have a maximum overlap at $D_{A}$ and that $W_{A}$ is positioned so that it reflects most of the two beams, except for their extreme Gaussian tails ($\sim10\sigma$).  Also, a removable beam stop has been placed in the path of the left-going photons near the source.  As stated above, when the left-going photons from the source are intercepted by such a beam stop, the non-coincident singles probabilities for Bob's detectors will be given by Eqns. \ref{eq:B1} and \ref{eq:B0}.  We wish to investigate the question of whether Bob will observe any change in the counting rates of his detectors that depends on whether the beam stop is in or out.

Naively it might appear that the new configuration would produce a large change in Bob's counting rates, because Alice could choose a phase $\phi_A$ for which the left-going wave components arriving at $D_{A}$ would interfere destructively and vanish or would interfere constructively and produce a maximum.  Arguments along these lines have been advanced by Anwar Sheikh\cite{Sh08} to justify a clever (but flawed)  one-photon faster-than-light communication scheme.  However, such expectations cannot be true, because they would violate quantum unitarity and the requirement that any left-going photon must be detected \textit{somewhere} with 100\% probability.  Unitarity (or equivalently, energy conservation) requires that any wave-mixing device that produces destructive interference in some locations must produce a precisely equal amount of constructive interference in other locations. The  $45^\circ$ wedge beam-combiner is no exception.

The flaw in such cancellation arguments is that in the previous examples we have always dealt with configurations in which only a single spatial mode of the photon is present.  In such cases, superposition can be used without considering wave trajectories, since the wave front for any given path arrives at a detector with a constant overall phase.  In the present configuration, the spatial profiles of the waves on Alice's two paths are truncated at the apex of the wedge mirror, producing non-Gaussian spatial modes,  and also must propagate in slightly different directions in order to overlap at the detector so they are definitely in different spatial modes.  Therefore, the phase of arriving waves is not constant and will depend on the location on the detector face.  Consequently, simple one-mode position-independent superposition cannot be used.

Instead, in order to calculate the differential probability of detection at a specific location on the face of detector $D_A$, one must propagate the waves from the wedge to the detector by doing a path integral of Huygens wavelets originating across the effective aperture of the wedge.  To get the overall detection probability, one must then integrate over locations on the detector face.  And since there are two quantum-distinguishable amplitudes arriving at the detector face, these must be converted to probabilities separately and then added.

The analysis of the wedge system is therefore much more challenging that those of the previous examples.  While analytic expressions can be obtained for the differential probability of two-particle detection with one of Bob's detectors and at some specific lateral position on $D_A$, the integration of that differential probability, a highly oscillatory function, over the face of $D_A$ cannot be done analytically.  Thus the analysis cannot produce equations predicting Bob's singles counts that can be directly compared with Eqns. \ref{eq:B1} and \ref{eq:B0} for the signal test.  Instead one must subtract the results of numerical integration from evaluations of Eqns. \ref{eq:B1} and \ref{eq:B0} using the same values for $\alpha$, $\phi_A$, and $\phi_B$ used in the numerical integration, and observe how close to zero is the calculated difference (which represents the potential nonlocal signal).

\begin{figure}
\includegraphics[width=15 cm]{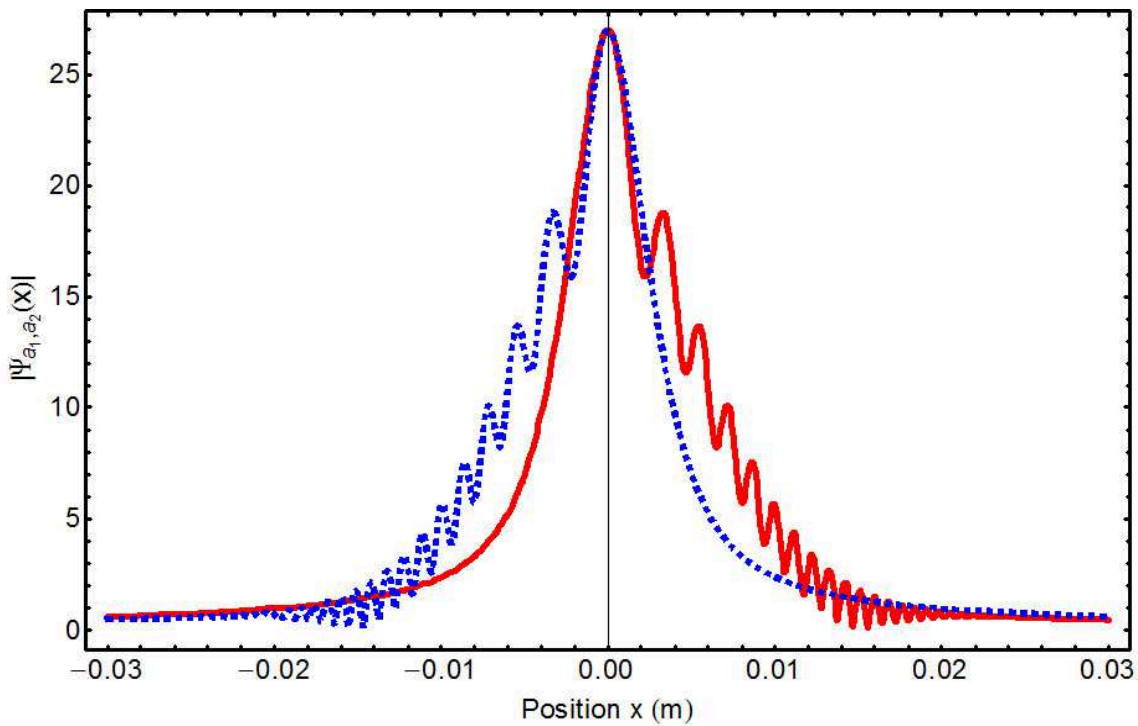}
\caption{(color online) Magnitudes of the wave functions $\Psi_{a1}$ (red/solid) and $\Psi_{a2}$ (blue/dotted) as functions of position $x$ on the face of detector $D_{A}$.  Oscillations are the result of Gaussian tail truncation by the apex of wedge mirror $W_{A}$.
}
\label{fig:6}       
\end{figure}

We have performed this analysis\cite{Ma14d} of the experiment shown in Fig. \ref{fig:5}, tweaking the mirror angles for maximum overlap of the waves on the two paths to detector $D_{A}$.  The calculation gives large analytical expressions for joint detection probability as a function of position on detector $D_{A}$, but these must be integrated numerically to obtain the position-independent probabilities.  Here Fig. \ref{fig:6} shows the overlap of the magnitudes of the wave functions for paths $a_1$ and $a_2$ vs. position.  The wave functions have a basic Gaussian profile with oscillations arising from the truncation of one Gaussian tail by $W_{A}$.
 
\begin{figure}
\includegraphics[width=15 cm]{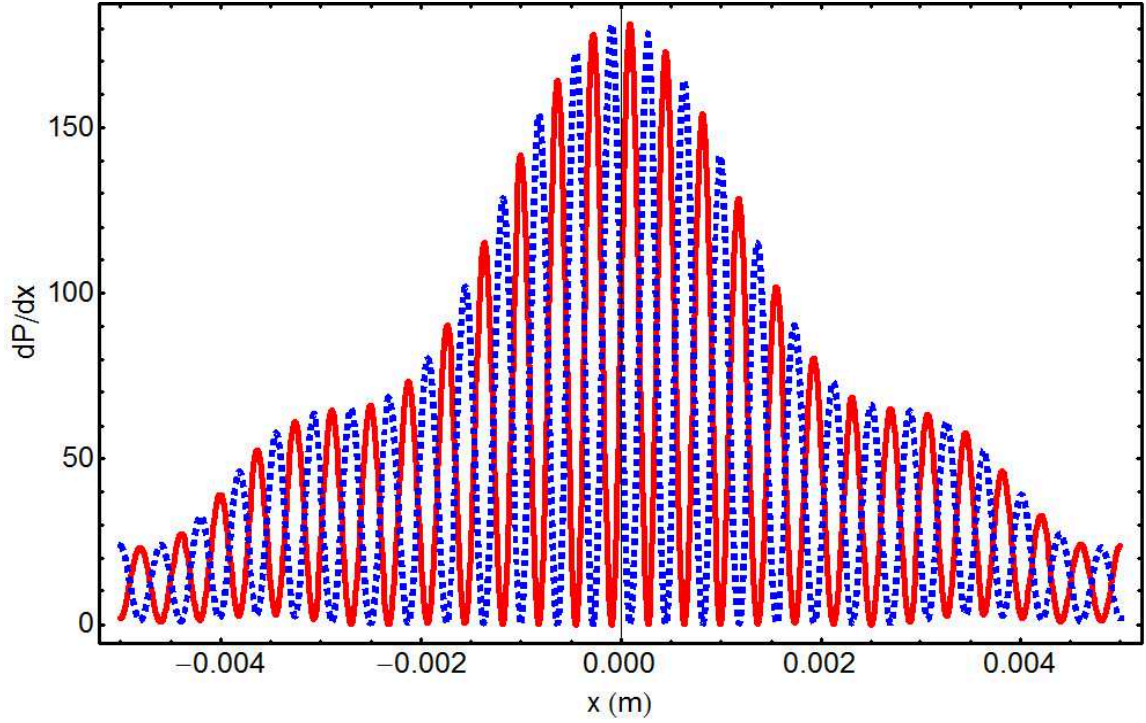}
\caption{(color online) Probabilities of coincident detections at $D_{A}$ and $D_{B_1}$ (red/solid) and at $D_{A}$ and $D_{B0}$ (blue/dotted) with $\alpha=0$, $\phi_A=0$, and $\phi_B=0$.
}
\label{fig:7}       
\end{figure}

Fig. \ref{fig:7} shows the corresponding probabilities for $\alpha=0$ (e.g., fully entangled) of coincident photon pairs at Alice's detector $D_A$ and at Bob's detectors $D_{B_1}$ and $D_{B_0}$. The probabilities are highly oscillatory because of the interference of the two waves and the phase walk of the wave functions with angle, analogous to two-slit interference.

\begin{figure}
\includegraphics[width=15 cm]{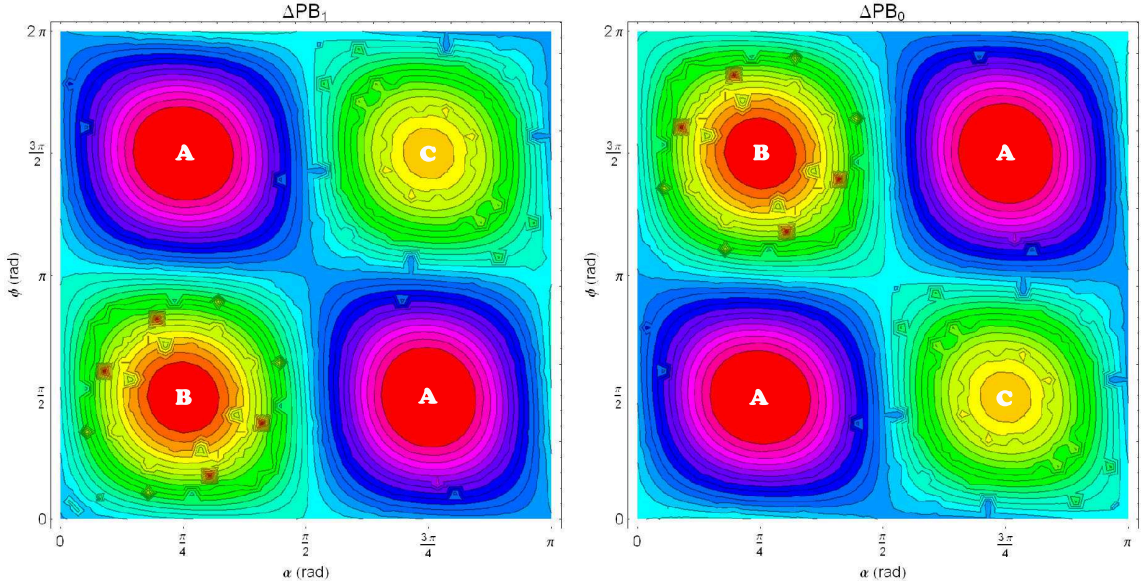}
\caption{(color online) Difference between numerical singles probabilities and evaluations of Eqns. \ref{eq:B1} and \ref{eq:B0}.  Here the regions labeled ``\textbf{A}" reach minima of $5.7 \times 10^{-7}$, the regions labeled  ``\textbf{B}" reach maxima of $6.08 \times 10^{-6}$, and the regions labeled  ``\textbf{C}" reach maxima of $5.51 \times 10^{-6}$.  Small blotches indicate regions in which numerical integration has produced errors. }
\label{fig:8}       
\end{figure}

To test the possibility of a nonlocal signal, we must integrate these probabilities over the extent of the detector face and calculate difference functions from these results and similar evaluations of Eqns. \ref{eq:B1} and \ref{eq:B0}.  We can expect some errors in numerical integration due to the oscillation shown in Fig. \ref{fig:7}.  The difference functions as 2-D contour plots in $\phi_B$ vs. $\alpha$ are shown in Fig. \ref{fig:8}.
 
Thus, the differences between the probabilities predicted by of Eqns. \ref{eq:B1} and \ref{eq:B0} and the numerically-integrated probabilities of Fig. \ref{fig:7} are on the order of a few parts per million.  This is equivalent to saying that they are the same, and that no nonlocal signal is possible using the wedge-modified configuration of Fig. \ref{fig:5}.

\section{Conclusions}
\label{sec:7}

We have investigated the possibility of nonlocal quantum signaling by analyzing polarization-entangled and path-entangled systems.  The conjecture\cite{Cr09} that nonlocal signaling might be possible by adjusting the entanglement to 71\% to permit coherence has proved to be incorrect.  Instead, we find that in all cases investigated the interference is blocked by a ``signal" interference pattern and an ``anti-signal" interference pattern that mask any observable interference when they are added, even when entanglement and coherence are simultaneously present.  This behavior is again attributed to what has been called the complementarity of one-particle and two-particle interference\cite{Ja93}.

Our conclusion, based on the standard formalism of quantum mechanics as applied to these \textit{gedankenexperiments}, is that no nonlocal signal can be transmitted from Alice to Bob by varying Alice's configuration in any of the ways discussed here.  In all of the cases studied, there are two quantum-distinguishable modes of entangled photon-pair behavior that each contain a``switch-able" interference pattern, but when these modes are superimposed, the two interference patterns always complement each other and together become invisible.  This is the mechanism by which the formalism of quantum mechanics blocks nonlocal signaling.  In the context of the standard quantum formalism, Nature appears to be well protected from the possibility of nonlocal signaling.  

\begin{acknowledgements}
This work was supported in part by the U. S. Department of Energy Office of Scientific Research.  We are grateful to Prof. Anton Zeilinger, Dr.~Radek Lapkiewicz, Prof.~Gerald Miller, Prof.~Yahuna Shih, and Prof.~James F.~Woodward for valuable comments, suggestions, and criticisms during the course of this work.
\end{acknowledgements}



\end{document}